\documentclass[aps,prl,reprint,superscriptaddress]{revtex4-2}
\usepackage{comment}
\usepackage{hyperref}
\usepackage{graphicx}%
\usepackage{amsmath,amssymb,amsfonts}%
\usepackage{ragged2e}
\usepackage{lineno}
\usepackage[title]{appendix}%
\usepackage{xcolor}%
\usepackage[justification=justified,singlelinecheck=false]{caption}
\begin{document}

\title{On the Mathematics of Information-Thermodynamics}


\author{Dallin Fisher}\email{djfishe7@asu.edu}

\author{Qi-Jun Hong}\email{qhong@alumni.caltech.edu}
\affiliation{Materials Science and Engineering, Arizona State University, Tempe, Arizona 85285, USA}



\begin{abstract}
We present a validation of the \texttt{asdf} method, an information-theoretic framework for computing thermodynamic entropy from molecular configurations. 
The method reformulates entropy estimation as the Shannon entropy of a residual mapping distribution defined between two decorrelated microstates. 
We demonstrate analytically that for the closed-form Hamiltonians with known solutions, the classical ideal gas and the one-dimensional harmonic oscillator's entropy obtained from the compressibility of the residual mapping object $\Delta$ reproduces the exact thermodynamic entropy. 
In each case, the conditional entropy of $\Delta$ with respect to an uncorrelated microstate is shown to coincide with the ensemble entropy derived from the canonical partition function. 
These results establish consistency between the \texttt{asdf} formalism and classical statistical mechanics for analytically solvable systems. We further discuss how the framework generalizes to interacting Hamiltonians. 
The analysis supports the interpretation of thermodynamic entropy as an information measure encoded geometrically in inter-microstate mappings and motivates application of the method to complex condensed phases.
\end{abstract}

\maketitle

\section{Introduction}

Entropy is a thermodynamic quantity that is notoriously difficult to compute in interacting systems. Closed-form solutions are rare, and it is common practice to validate any new entropy method against systems with analytic entropies, then evaluate how the method generalizes beyond these solvable limits. In this section we do exactly this for two canonical benchmarks: the classical ideal gas and the classical one-dimensional (1-D) harmonic oscillator. We derive their entropies (i) in the standard canonical-ensemble framework via the partition function and (ii) via an information-theoretic construction based on the Shannon entropy\cite{shannon} defined as

\begin{equation}
\label{eq:shannon}
H(X)=-\sum_{i}{p(x_i)\log{p(x_i)}},
\end{equation}

applied to the minimum mapping vectors (residuals) between two sufficiently decorrelated microstates. The goal is to make explicit how the two perspectives are mathematically consistent, not to replace the standard derivations. In later sections, we will demonstrate where the mapping-vector construction may remain practical when analytic integration is not.

In classical statistical mechanics, the dynamics and energetics of an ensemble are encoded by a Hamiltonian defined on phase space $(\mathbf{q},\mathbf{p})$, where $\mathbf{q}$ and $\mathbf{p}$ denote conjugate coordinates and momenta. For a broad class of systems relevant here, the Hamiltonian has the form
\[
H(\mathbf{p},\mathbf{q})
=
\sum_{i=1}^{N}\frac{\mathbf{p}_i^2}{2m}
+
V(\mathbf{q}),
\label{eq:H_general}
\]
i.e., it is quadratic in momenta and all nontrivial structure resides in the potential $V(\mathbf{q})$. In the canonical ensemble, the Boltzmann weight\cite{boltzmann} of a microstate is
\begin{align}
e^{-\beta H(\mathbf{p},\mathbf{q})}
&=
\exp\!\left[-\beta\left(\sum_{i=1}^N\frac{\mathbf{p}_i^2}{2m}+V(\mathbf{q})\right)\right]
\nonumber\\
&=
\exp\!\left(-\beta \sum_{i=1}^N\frac{\mathbf{p}_i^2}{2m}\right)\,
\exp\!\left(-\beta V(\mathbf{q})\right),
\label{eq:boltz_factorization}
\end{align}
where $\beta=(k_BT)^{-1}$. When $V(\mathbf{q})$ does not depend on $\mathbf{p}$, this factorization implies that the partition function factorizes into a momentum and configurational contribution. The canonical partition function for $N$ identical classical particles is
\begin{equation}
Z(N,V,T)
=
\frac{1}{N!\,h^{3N}}
\int d^{3N}\mathbf{q}\int d^{3N}\mathbf{p}\;
e^{-\beta H(\mathbf{p},\mathbf{q})}.
\label{eq:Z_general}
\end{equation}
Here $N!$ accounts for indistinguishability and $h^{3N}$ is the conventional phase-space cell that renders $Z$ dimensionless and fixes the additive constant of the absolute entropy (equivalently: it encodes the resolution scale at which microstates are counted). The thermodynamic functions follow from $Z$, entropy being in the same fashion as Gibb's \cite{Gibbs2011-qz}:
\begin{equation}
E = -\frac{\partial}{\partial \beta}\ln Z,
\qquad
S = k_B\left(\ln Z + \beta E\right).
\label{eq:S_from_Z}
\end{equation}

It is this standard approach that we will use to construct analytic solutions for the thermodynamic entropy of systems with idealized Hamiltonians. We will then verify the \texttt{asdf} method reproduces these analytic results, before discussing the generalization of the \texttt{asdf} framework for Hamiltonians of real physical systems, including a comparison against already existing paradigms. 

Before we begin with the ideal gas, it is important to define the \texttt{asdf} method in this context. \texttt{asdf} asserts that there exists an object $\Delta$ that when compressed optimally, is related to the thermodynamic entropy of a microstate $Y$ as

\begin{equation}\label{eq:landauers}
    S(Y)=k_B\ln{2}\,\mathrm{H}(\Delta\mid X)
\end{equation}

where H measures the shannon entropy in bits of the argument and $X$ is a sufficiently uncorrelated microstate in the ensemble. Eq. \eqref{eq:landauers} is essentially Landauer's principle\cite{LaundauersPrinciple}. While it is true $S(Y)\propto \mathrm{H}(Y)$, there are advantages, as will be discussed, in defining $\Delta$ as the minimum distance map from an uncorrelated microstate $X$ to $Y$:
\[
\Delta(X,Y)
=
\left\{\, y_{\sigma(k)} - x_k \,\right\}_{k=1}^{N}
\]

\[
\sigma(k)
=
\operatorname*{arg\,min}_{j \in \{1,\dots,N\}}
\|\, y_j - x_k \,\|
\]

\[
\text{so that}\qquad
y_{\sigma(k)} = x_k + \delta_k, \quad \delta_k \in \Delta(X,Y).
\]

Essentially, $\Delta$ is a one to many map from $X$ to $Y$ so that $\Delta$ is pointing from a member of $X$ to a member(s) of $Y$. We then assert
\begin{equation}\label{eq:asdf_entropy}
\mathrm{H}(Y)=\mathrm{H}(\Delta\mid X)
\end{equation}
as illustrating the entropy of the microstate $Y$ can be obtained as the entropy of the mapping $\Delta$ given another microstate $X$. In other words, if I know how $X$ points to $Y$, then I can construct $Y$ given $X$. To test whether Eq. \eqref{eq:asdf_entropy} holds for thermodynamic ensembles producing microstates $(X,Y)$, we construct the object $\Delta$ for the system and compute it's conditional entropy. We begin with doing so for the ideal gas.

\section{Ideal Gas}

For a monatomic ideal gas, the Hamiltonian contains only kinetic energy,
\begin{equation}
H_{\mathrm{ig}}(\mathbf{p}) = \sum_{i=1}^{N}\frac{\mathbf{p}_i^2}{2m},
\label{eq:H_ideal_gas}
\end{equation}
so the configurational dependence disappears. 
Indeed the configurational dependence being nonexistent, along with the Hamiltonian being quadratic in momentum, is exactly what makes this system the most idealized. 
Since the momentum term is identical for all real Hamiltonians, the boltzmann factor is always gaussian in momentum; the momentum term is then always the same and will be recalled when solving the momentum part of any generalized Hamiltonian. 

Substituting \eqref{eq:H_ideal_gas} into \eqref{eq:Z_general} yields
\begin{align}
Z_{\mathrm{ig}}
&=
\frac{1}{N!\,h^{3N}}
\int d^{3N}\mathbf{q}\int d^{3N}\mathbf{p}\;
\exp\!\left(-\beta\sum_{i=1}^{N}\frac{\mathbf{p}_i^2}{2m}\right)
\nonumber\\
&=
\frac{1}{N!\,h^{3N}}
\left(\int d^{3N}\mathbf{q}\right)
\left(\int d^{3N}\mathbf{p}\;
\exp\!\left(-\beta\sum_{i=1}^{N}\frac{\mathbf{p}_i^2}{2m}\right)\right).
\label{eq:Z_ig_factorized}
\end{align}
The configurational integral is simply the volume of configuration space,
\begin{equation}
\int d^{3N}\mathbf{q} = V^N.
\label{eq:config_integral_ig}
\end{equation}
The momentum integral is Gaussian. First note the separability across particles and Cartesian components:
\begin{align}
\int d^{3N}\mathbf{p}\;
&\exp\!\left(-\beta\sum_{i=1}^{N}\frac{\mathbf{p}_i^2}{2m}\right)
\nonumber\\
&=
\prod_{i=1}^{N}\left(\int_{\mathbb{R}^3} d^3\mathbf{p}_i\;
\exp\!\left(-\beta\frac{\mathbf{p}_i^2}{2m}\right)\right)
\nonumber\\
&=
\left(\int_{\mathbb{R}^3} d^3\mathbf{p}\;
e^{-\beta \mathbf{p}^2/(2m)}\right)^N
\nonumber\\
&=
\left(\prod_{\alpha=x,y,z}\int_{-\infty}^{\infty} dp_\alpha\; e^{-\beta p_\alpha^2/(2m)}\right)^N.
\label{eq:momentum_integral_factorization}
\end{align}
Using the standard Gaussian integral $\int_{-\infty}^\infty e^{-a p^2}dp = \sqrt{\pi/a}$, with $a=\beta/(2m)$,
\begin{equation}
\int_{-\infty}^{\infty} dp_\alpha\; e^{-\beta p_\alpha^2/(2m)} = \sqrt{\frac{2\pi m}{\beta}}.
\label{eq:1d_gaussian}
\end{equation}
Therefore,
\[
\int_{\mathbb{R}^3} d^3\mathbf{p}\; e^{-\beta \mathbf{p}^2/(2m)}
=
\left(\sqrt{\frac{2\pi m}{\beta}}\right)^3
=
\left(2\pi m k_BT\right)^{3/2},
\label{eq:3d_gaussian}
\]
and hence
\begin{equation}
\int d^{3N}\mathbf{p}\;
\exp\!\left(-\beta\sum_{i=1}^{N}\frac{\mathbf{p}_i^2}{2m}\right)
=
\left(2\pi m k_BT\right)^{3N/2}.
\label{eq:3N_gaussian}
\end{equation}
Plugging \eqref{eq:config_integral_ig} and \eqref{eq:3N_gaussian} into \eqref{eq:Z_ig_factorized} gives
\begin{equation}
Z_{\mathrm{ig}}
=
\frac{V^N}{N!\,h^{3N}}\left(2\pi m k_BT\right)^{3N/2}.
\label{eq:Z_ig_preLambda}
\end{equation}
Define the thermal de Broglie wavelength
\[
\Lambda \equiv \frac{h}{\sqrt{2\pi m k_BT}},
\label{eq:Lambda_def}
\]
so that
\[
\frac{\left(2\pi m k_BT\right)^{3/2}}{h^3} = \frac{1}{\Lambda^3}.
\label{eq:Lambda_identity}
\]
Then \eqref{eq:Z_ig_preLambda} becomes
\[
Z_{\mathrm{ig}}
=
\frac{1}{N!}\left(\frac{V}{\Lambda^3}\right)^N.
\label{eq:Z_ig_final}
\]

To obtain the entropy, we compute $\ln Z_{\mathrm{ig}}$:
\[
\ln Z_{\mathrm{ig}}
=
N\ln V - \ln N! - 3N\ln \Lambda.
\label{eq:lnZ_ig}
\]
Stirling's approximation $\ln N!\approx N\ln N - N$ yields
\begin{align}
\ln Z_{\mathrm{ig}}
&\approx N\ln V - (N\ln N - N) - 3N\ln\Lambda
\nonumber\\
&=
N\ln\!\left(\frac{V}{N}\right) + N - 3N\ln\Lambda
\nonumber\\
&=
N\left[\ln\!\left(\frac{V}{N\Lambda^3}\right)+1\right].
\label{eq:lnZ_ig_stirling}
\end{align}
Next compute the energy via \eqref{eq:S_from_Z}. One may either use equipartition or differentiate $\ln Z_{\mathrm{ig}}$ with respect to $\beta$; the result is
\[
E_{\mathrm{ig}} = \frac{3}{2}Nk_BT,
\qquad
\beta E_{\mathrm{ig}}=\frac{3}{2}N.
\label{eq:E_ig}
\]
Finally, \eqref{eq:S_from_Z} gives
\begin{align}
\frac{S_{\mathrm{ig}}}{k_B}
&=
\ln Z_{\mathrm{ig}} + \beta E_{\mathrm{ig}}
\approx
N\left[\ln\!\left(\frac{V}{N\Lambda^3}\right)+1\right]+\frac{3}{2}N
\nonumber\\
&=
N\left[\ln\!\left(\frac{V}{N\Lambda^3}\right)+\frac{5}{2}\right].
\label{eq:Sackur_Tetrode}
\end{align}
Thus, per particle,
\begin{equation}
\frac{S_{\mathrm{ig}}}{Nk_B}
=
\ln\!\left(\frac{V}{N\Lambda^3}\right)+\frac{5}{2}.
\label{eq:S_ig_per_particle}
\end{equation}
It is often conceptually helpful to separate this into a configurational piece and a momentum piece:
\begin{equation}
\frac{S_{\mathrm{ig}}}{Nk_B}
=
\underbrace{\left[\ln\!\left(\frac{V}{N}\right)+1\right]}_{\text{configurational}}
+
\underbrace{\left[\ln\!\left(\frac{1}{\Lambda^3}\right)+\frac{3}{2}\right]}_{\text{momentum}}.
\label{eq:S_ig_split}
\end{equation}
The momentum contribution is universal for any ensemble whose kinetic energy is quadratic in $\mathbf{p}$; the configurational contribution is where nontrivial structure enters through $V(\mathbf{q})$.

\vspace{0.75em}

We now reproduce the \emph{configurational} contribution to the ideal-gas entropy by computing the Shannon entropy of the minimum mapping (nearest-neighbor) residuals between two independent microstates.

Model each microstate as a homogeneous Poisson point process (PPP) in $\mathbb{R}^3$ with intensity (number density) $\lambda = N/V$. Let $X$ and $Y$ be two independent PPP realizations of the same intensity $\lambda$. For a fixed point $X_i\in X$, define the nearest-neighbor distance to the process $Y$:
\[
R \equiv \min_{Y_j\in Y}\|X_i-Y_j\|.
\]
The event $\{R>r\}$ means that the ball of radius $r$ centered at $X_i$ contains no points of $Y$. For a PPP, the number of points in a region of volume $b(r)$ is Poisson-distributed with mean $\lambda b(r)$, hence the void probability is
\[
\mathbb{P}(R>r)=\exp\!\big(-\lambda\, b(r)\big),
\qquad
b(r)=\frac{4}{3}\pi r^3.
\]
Therefore the cumulative distribution function (CDF) of $R$ is
\begin{equation}\label{eq:FR}
F_R(r)=\mathbb{P}(R\le r)=1-\exp\!\left(-\lambda \frac{4}{3}\pi r^3\right).
\end{equation}
Define
\[
a \equiv \lambda\,\frac{4}{3}\pi.
\]
Differentiating the CDF gives the radial probability density function (PDF) of the nearest-neighbor distance:
\[
f_R(r)=\frac{dF_R}{dr}
=
4\pi \lambda r^2 \exp(-a r^3),
\qquad r\ge 0.
\]
It is convenient to define the corresponding \emph{isotropic 3D residual-vector density} $g(r)$ (a density per unit volume in $\mathbb{R}^3$) through the standard relation
\begin{align}
f_R(r) &= 4\pi r^2\, g(r),
\nonumber\\
&\Longrightarrow\qquad
g(r)=\frac{f_R(r)}{4\pi r^2}
=\lambda\,\exp(-a r^3).
\label{eq:g_r_def}
\end{align}

We define the configurational mapping entropy as the Shannon entropy of the residual-vector distribution,
\[
h_{\mathrm{map}}
\equiv
-\int_{\mathbb{R}^3} g(\mathbf{r})\ln g(\mathbf{r})\,d^3\mathbf{r}.
\]
By isotropy, $g(\mathbf{r})=g(r)$ and $d^3\mathbf{r}=4\pi r^2\,dr$, so
\begin{align}
h_{\mathrm{map}}
&=
-\int_{0}^{\infty} g(r)\ln g(r)\;4\pi r^2\,dr
\nonumber\\
&=
-\int_{0}^{\infty} f_R(r)\,\ln g(r)\,dr
\nonumber\\
&=
-\mathbb{E}\big[\ln g(R)\big].
\label{eq:hmap_expect}
\end{align}
From \eqref{eq:g_r_def},
\[
\ln g(r) = \ln\lambda - a r^3,
\]
so \eqref{eq:hmap_expect} becomes
\begin{equation}
h_{\mathrm{map}}
=
-\mathbb{E}[\ln\lambda] + a\,\mathbb{E}[R^3]
=
-\ln\lambda + a\,\mathbb{E}[R^3].
\label{eq:hmap_two_terms}
\end{equation}

It remains to compute $\mathbb{E}[R^3]$ using $f_R(r)$:
\begin{align}
\mathbb{E}[R^3]
&=
\int_{0}^{\infty} r^3 f_R(r)\,dr
\nonumber\\
&=
4\pi\lambda\int_{0}^{\infty} r^5 e^{-a r^3}\,dr
\nonumber\\
&=
\frac{4\pi\lambda}{3}a^{-2}
\end{align}
Using $a=\lambda\frac{4}{3}\pi$, this simplifies immediately to
\[
\mathbb{E}[R^3]=\frac{1}{a}.
\]
Substituting into \eqref{eq:hmap_two_terms} gives
\[
h_{\mathrm{map}}
=
-\ln\lambda + a\left(\frac{1}{a}\right)
=
\ln\!\left(\frac{V}{N}\right)+1.
\label{eq:hmap_final}
\]
Thus the configurational entropy is
\[
S_{\mathrm{config}}
=
Nk_B\,h_{\mathrm{map}}
=
Nk_B\left[\ln\!\left(\frac{V}{N}\right)+1\right].
\label{eq:Sconfig_from_mapping}
\]
Adding the universal momentum contribution (\ref{eq:S_ig_split}) for a quadratic kinetic term,
\[
\frac{S_{\mathrm{mom}}}{Nk_B}
=
\ln\!\left(\frac{1}{\Lambda^3}\right)+\frac{3}{2}
\]
yields the total ideal-gas entropy,
\begin{equation}\label{eq:ig_from_shannon}
\frac{S_{\mathrm{ig}}}{Nk_B}
=
\ln\!\left(\frac{V}{N\Lambda^3}\right)+\frac{5}{2}.
\end{equation}

Equation \ref{eq:ig_from_shannon} is identical to equation \ref{eq:S_ig_per_particle}.

\begin{figure}
    \centering
    \includegraphics[width=1.0\linewidth]{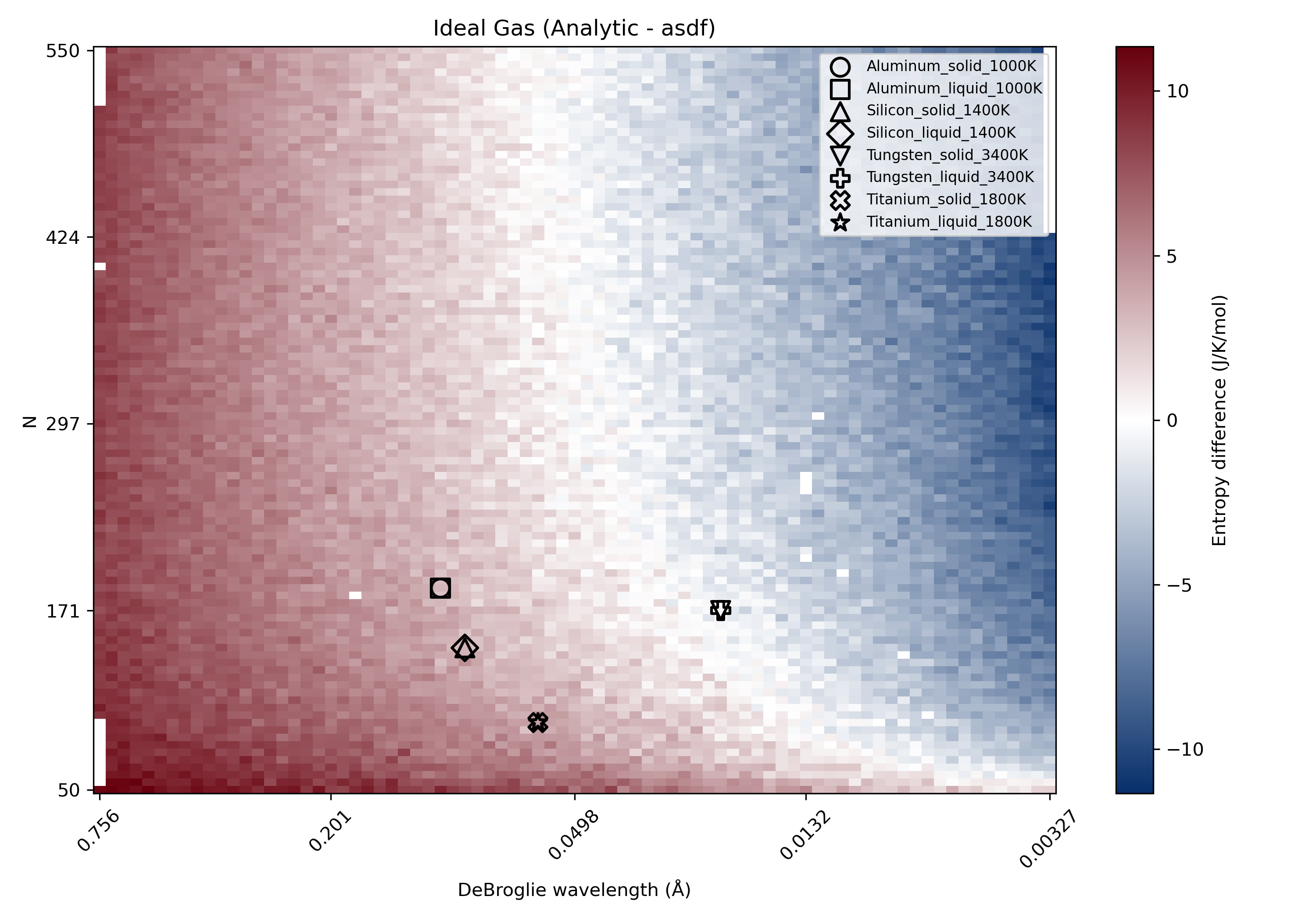}
    \caption{Heatmap showing the discrepancy between the \texttt{asdf} method and the analytic solutions for entropy of an ideal gas for $~8,000$ ensembles with varying $N$ and $\Lambda$. For real systems simulated with sufficiently sized cells, \texttt{asdf} is in good agreement with the analytic solutions.}
    \label{fig:ideal_gas}
\end{figure}

In addition to the analytic derivation above, it is important to verify that \texttt{asdf} numerically implements the intended quantity, namely the conditional mapping entropy defined in Eq. \eqref{eq:asdf_entropy}. To do so, we construct synthetic ensembles representing an ideal gas, for which the entropy is already known analytically from Eq. \eqref{eq:S_ig_per_particle}, depending only on $N$, $V$, and the thermal de Broglie wavelength $\Lambda$. Each ensemble is generated by uniformly and independently distributing $N$ atoms throughout a fixed volume 
$V$, with no explicit momentum assigned. Repeating this procedure yields a statistical ensemble of ideal-gas microstates. For a chosen $\Lambda$, corresponding to a particular mass and temperature, the analytic entropy is evaluated from Eq. \eqref{eq:S_ig_per_particle}, and the same ensemble is then provided to \texttt{asdf}, which compresses the mapping data and returns an entropy estimate obtained statistically from the ensemble. By repeating this procedure over many ensembles while varying $N$ and $\Lambda$ we directly test whether the numerical implementation converges to the known ideal-gas result. The comparison is shown in Fig. \ref{fig:ideal_gas}.

The results exhibit the expected trends. First, the accuracy of \texttt{asdf} improves with increasing $N$. For small systems, the finite number of particles does not adequately sample the ideal-gas distribution, and the entropy is correspondingly underestimated. As $N$ increases, the sampled microstates more faithfully represent the bulk distribution, and the \texttt{asdf} value approaches the analytic expression asymptotically. Second, the agreement also depends on the Broglie length scale $\Lambda$. When $\Lambda$ is chosen too small relative to $V$, numerical inaccuracies within \texttt{asdf} begin to dominate and artificially inflate the estimated entropy value. By contrast, when $\Lambda$ is too large relative to $V$, the discretization becomes overly coarse and the finite ensemble no longer contains enough recoverable information, leading to an underestimation of the entropy. Thus, the ideal-gas test confirms the expected large-$N$ convergence of \texttt{asdf}, and suggests accuracy relies on $V$ and $\Lambda$ being related such that $\Lambda$ optimally discretizes $V$.

Alongside the ideal-gas results, we superimpose data from single-element material systems computed in the main \texttt{asdf} study \cite{asdf}. Each system is plotted at its corresponding de Broglie wavelength and particle number. These material systems fall within a regime that exhibits relatively small bias, primarily due to their finite system sizes. From the ideal-gas analysis, this regime is expected to produce a slight underestimation of the absolute thermodynamic entropy. However, because phase comparisons are performed at fixed temperature and identical system size, this bias is effectively systematic and cancels when taking entropy differences. As a result, no additional error is introduced into $\Delta S$ measurements due to deviations from the ideal-gas limit.

Overall, for finite and moderately sized systems, discrepancies between the analytic and numerical ideal-gas results can lead to a small downward bias in absolute entropy. Nevertheless, the \texttt{asdf} method performs reliably in capturing thermodynamic entropy at physically relevant temperatures and masses, particularly when system sizes are sufficiently large and the compression threshold is chosen consistently with the underlying physical length scales.

\section{One-Dimensional Harmonic Oscillator}
\label{sec:classical_ho}
We now consider the 1-D harmonic oscillator, whose Hamiltonian is quadratic in both position and momentum:
\begin{equation}
H_{\mathrm{ho}}(q,p)
=
\frac{p^2}{2m}
+
\frac{1}{2}m\omega^2 q^2
\label{eq:H_HO}
\end{equation}
The one-dimensional harmonic oscillator is an idealized system where the momentum and position are quadratic. For the ideal gas, only the momentum is quadratic, and the position term did not exist. This is naturally an extension of the ideal gas in which we introduce a symmetric position term so the bolzmann factor is gaussian in position and momentum. It is no coincidence that both the position and the momentum are solved nearly identically to the ideal gas momentum case.

The classical canonical partition function is
\[
Z_{\mathrm{ho}}
=
\frac{1}{h}
\int_{-\infty}^{\infty}\int_{-\infty}^{\infty}
\exp\!\left[-\beta\left(\frac{p^2}{2m}+\frac{1}{2}m\omega^2 q^2\right)\right]\,dq\,dp.
\label{eq:Z_HO_def}
\]
Factorizing the integrals,
\[
Z_{\mathrm{ho}}
=
\frac{1}{h}
\left(\int_{-\infty}^{\infty} e^{-\beta p^2/(2m)}\,dp\right)
\left(\int_{-\infty}^{\infty} e^{-\beta m\omega^2 q^2/2}\,dq\right).
\label{eq:Z_HO_factored}
\]
Using \eqref{eq:1d_gaussian} for the momentum integral and again $\int e^{-a q^2}dq=\sqrt{\pi/a}$ with $a=\beta m\omega^2/2$ for the configurational integral,
\[
\int_{-\infty}^{\infty} e^{-\beta m\omega^2 q^2/2}\,dq
=
\sqrt{\frac{2\pi}{\beta m\omega^2}}.
\label{eq:q_gaussian}
\]
Therefore,
\[
Z_{\mathrm{ho}}
=
\frac{1}{h}
\sqrt{\frac{2\pi m}{\beta}}
\sqrt{\frac{2\pi}{\beta m\omega^2}}
=
\frac{1}{h}\,\frac{2\pi}{\beta\omega}
=
\frac{2\pi k_BT}{h\omega}.
\label{eq:Z_HO_final}
\]
Compute energy:
\begin{align*}
\ln Z_{\mathrm{ho}} 
&= 
\ln\!\left(\frac{2\pi}{h\omega}\right) - \ln\beta
\nonumber\\
&\quad\Rightarrow\quad
E_{\mathrm{ho}} = -\frac{\partial}{\partial \beta}\ln Z_{\mathrm{ho}} = \frac{1}{\beta}=k_BT,
\label{eq:E_HO}
\end{align*}
so $\beta E_{\mathrm{ho}}=1$. Hence the entropy is
\begin{equation}
\frac{S_{\mathrm{ho}}}{k_B}
=
\ln Z_{\mathrm{ho}} + \beta E_{\mathrm{ho}}
=
1+\ln\!\left(\frac{2\pi k_BT}{h\omega}\right).
\label{eq:S_HO}
\end{equation}

\vspace{0.75em}
Because \eqref{eq:H_HO} is quadratic, the canonical distribution is Gaussian in both $q$ and $p$:
\[
q\sim \mathcal{N}(0,\sigma_q^2),
\qquad
\sigma_q^2=\frac{k_BT}{m\omega^2},
\label{eq:q_dist}
\]
\[
p\sim \mathcal{N}(0,\sigma_p^2),
\qquad
\sigma_p^2=m k_BT,
\label{eq:p_dist}
\]
and $q$ and $p$ are independent. Consider two independent microstates $(q_1,p_1)$ and $(q_2,p_2)$ drawn from this equilibrium distribution. Define residuals
\[
\Delta q = q_2-q_1,
\qquad
\Delta p = p_2-p_1.
\label{eq:Delta_defs}
\]
Differences of independent Gaussians remain Gaussian with variances adding:
\[
\Delta q \sim \mathcal{N}(0,2\sigma_q^2),
\qquad
\Delta p \sim \mathcal{N}(0,2\sigma_p^2).
\label{eq:Delta_dists}
\]
The differential entropy of $\mathcal{N}(0,\sigma^2)$ is
\[
h\!\left(\mathcal{N}(0,\sigma^2)\right)
=
\frac{1}{2}\ln\!\left(2\pi e\,\sigma^2\right).
\label{eq:gauss_entropy}
\]
Since $\Delta q$ and $\Delta p$ are independent,
\begin{align*}
h(\Delta q,\Delta p)
&=
h(\Delta q)+h(\Delta p)
\nonumber\\
&=
\frac{1}{2}\ln\!\left(2\pi e\,(2\sigma_q^2)\right)
+
\frac{1}{2}\ln\!\left(2\pi e\,(2\sigma_p^2)\right)
\nonumber\\
&=
\ln\!\left(2\pi e\right)
+\frac{1}{2}\ln\!\left(4\sigma_q^2\sigma_p^2\right).
\label{eq:hDelta_expand}
\end{align*}
Using $\sigma_q^2\sigma_p^2 = (k_BT/\omega)^2$, we have
\[
h(\Delta q,\Delta p)
=
\ln\!\left(2\pi e\right)+\ln\!\left(2\frac{k_BT}{\omega}\right)
=
\ln\!\left(4\pi e\,\frac{k_BT}{\omega}\right).
\label{eq:hDelta_final}
\]
To compare with thermodynamic entropy, we must specify a phase-space resolution. Discretizing phase space into cells of area $\Delta q\,\Delta p$ shifts the entropy by $-\ln(\Delta q\,\Delta p)$. Taking the conventional choice $\Delta q\,\Delta p = h$ yields
\begin{align*}
\frac{S_{\mathrm{ho}}}{k_B}
=
h(q,p) - \ln h
&=
\left[h(\Delta q,\Delta p)\right] - \ln h
\nonumber\\
&=
1+\ln\!\left(\frac{4\pi k_BT}{h\omega}\right),
\label{eq:HO_match}
\end{align*}
which nearly matches \eqref{eq:S_HO}, save it for a factor of 2 in the logarithm. The discrepancy can be attributed to our calculation of the shannon entropy of the object $h(\Delta q, \Delta p)$, which does not correspond exactly to the entropy of a single microstate, that is $h(\Delta q, \Delta p) \neq h(q_1,p_1)$. Instead, as is utilized in the \texttt{asdf} method, the information contained in a microstate is equal to the information contained in the mapping from a \textit{given} microstate subject to macrostate constraints (NVE). Therefore if we define $\Delta=Y-X$ to be the map $(x,y)\leftrightarrow(x,\delta)$ representing the map from uncorrelated microstates $X$ to $Y$, then 
\[
h(Y)=h(\Delta\mid X).
\]
It is clear in the derivation we computed $h(\Delta)$ in which case using $\Delta\sim\mathcal{N}(0,\Sigma_{\Delta})$ and $Y\sim\mathcal{N}(0,\Sigma)$ with $\Sigma_{\Delta}=2\Sigma$ gives
\begin{align}\label{eq:s_delta}
    &h(\mathcal{N}(0,\Sigma))=\frac{1}{2}\ln{((2\pi e)^d \mathrm{det}\Sigma))}
    \nonumber\\
    &\rightarrow
    h(\Delta) - h(Y)=\frac{1}{2}\ln{\frac{\mathrm{det}(2\Sigma)}{\mathrm{det}\Sigma}}=\frac{d}{2}\ln{2}.
\end{align}
For the harmonic oscillator (but not in general), $\Delta$ is bijective; therefore we can use equation \ref{eq:s_delta} and $d=2$ corresponding to the 2 degrees of freedom, and hence $h(\Delta)$ picks up an extra $\ln{2}$. Therefore,
\begin{align*}
h(Y)
&=h(\Delta) - \ln{2}
\nonumber\\
&=1+\ln{\frac{2\pi k_B T}{h \omega}}
\end{align*}
which agrees with equation \ref{eq:S_HO}. 
One might question why it seemed the Shannon entropy of the map between microstates defined for the ideal gas was sufficient. This is because $F_R(r)$ defined in equation \ref{eq:FR} represents the probability an atom $y_j$ is found within a sphere of radius $r$ from an arbitrary point for which we have no locational information included. Therefore, it is clear the \texttt{asdf} method reproduces the analytic thermodynamic entropies for the ideal gas and harmonic oscilator. We now discuss generalizing the Hamiltonian to real systems.

\section{Entropy of Mixing}
For generalized Hamiltonians describing real materials, multiple atomic species may be present rather than a single component. Consequently, any method intended to compute thermodynamic entropy must account for the entropy associated with the mixing of different species.
We therefore consider a binary mixture on a fixed lattice with species \(A\) and \(B\), present at molar fractions \(x_A\) and \(x_B\), where
\[
x_A + x_B = 1.
\]

The entropy of mixing describes the entropy change that arises when species \(A\) and \(B\) are randomly distributed on lattice sites with no energetic preference for unlike or like neighbors, yielding
\begin{equation}
\label{eq:mixing_entropy}
S_{\mathrm{mix}} = -Nk_B\left(x_A \ln x_A + x_B \ln x_B\right).
\end{equation}
We will now show how eq. \ref{eq:mixing_entropy} can be derived from the usual conditional mapping between uncorrelated microstates. Let \(X\) denote a reference microstate and \(Y\) a sampled microstate of the same composition, recalling that \(H(\Delta \mid X) = H(Y \mid X) = H(Y).\) Define the sitewise mapping object \(\Delta_i\) by
\[
\Delta_i =
\begin{cases}
a, & X_i = Y_i \\
b, & X_i \neq Y_i
\end{cases}
\]
where \(a\) denotes ``same'' and \(b\) denotes ``different''.
A simple visualization of the mapping is shown in Table~\ref{tab:ideal_mixing_mapping_example}.

\begin{table}[h]
\centering
\caption{Example binary lattice mapping from a reference microstate \(X\) to a sampled microstate \(Y\), and the resulting mapping object \(\Delta\). Here \(a\) denotes sites for which \(X_i = Y_i\), and \(b\) denotes sites for which \(X_i \neq Y_i\).}
\label{tab:ideal_mixing_mapping_example}
\begin{tabular}{c c c c c}
\hline
Site \(i\) & 1 & 2 & 3 & 4 \\
\hline
\(X_i\) & \(A\) & \(B\) & \(A\) & \(B\) \\
\(Y_i\) & \(A\) & \(A\) & \(B\) & \(B\) \\
\(\Delta_i\) & \(a\) & \(b\) & \(b\) & \(a\) \\
\hline
\end{tabular}
\end{table}

For a site \(i\) with \(X_i = A\), the symbol \(a\) occurs when \(Y_i = A\), while \(b\) occurs when \(Y_i = B\). For an ideal mixture, these occur with probabilities
\[
P(a \mid X_i = A) = x_A,
\qquad
P(b \mid X_i = A) = x_B.
\]
Therefore,
\[
H(\Delta_i \mid X_i = A)
=
- x_A \ln x_A - x_B \ln x_B.
\]

Likewise, for a site \(i\) with \(X_i = B\), the symbol \(a\) occurs when \(Y_i = B\), while \(b\) occurs when \(Y_i = A\), so that
\[
P(a \mid X_i = B) = x_B,
\qquad
P(b \mid X_i = B) = x_A,
\]
and hence
\[
H(\Delta_i \mid X_i = B)
=
- x_B \ln x_B - x_A \ln x_A.
\]

Since these two expressions are equal, the conditional entropy \(H(\Delta_i \mid X_i)\) is their probability-weighted average:
\begin{align*}
H(\Delta_i \mid X_i)
&=
x_A H(\Delta_i \mid X_i = A)
+
x_B H(\Delta_i \mid X_i = B) \\
&=
(x_A + x_B)\left(- x_A \ln x_A - x_B \ln x_B\right) \\
&=
- x_A \ln x_A - x_B \ln x_B.
\end{align*}

Thus, the sitewise conditional mapping entropy is exactly the mixing entropy per site,
\[
H(\Delta_i \mid X_i)
=
- x_A \ln x_A - x_B \ln x_B.
\]
Summing over \(N\) lattice sites gives
\[
H(\Delta \mid X)
=
- N \left( x_A \ln x_A + x_B \ln x_B \right),
\]
and therefore
\[
S_{\mathrm{mix}}
=
k_B H(\Delta \mid X)
=
- N k_B \left( x_A \ln x_A + x_B \ln x_B \right),
\]
which is the standard entropy of mixing as seen in eq. \eqref{eq:mixing_entropy}.
From this it is clear that when multicomponent materials are considered, the mapping object must encode the species information for each mapped site, specifying the species of \(Y_i\) relative to the corresponding element of \(X\).

To emphasize the importance of the conditional relation \(H(Y)=H(\Delta\mid X)\), consider instead the marginal entropy \(H(\Delta)\). The probability of observing the symbol \(b\) is now
\begin{equation*}
P(b) = P(X_i \neq Y_i) = x_A x_B + x_B x_A = 2 x_A x_B,
\end{equation*}
so that
\begin{equation*}
H(\Delta_i)
=
-(1-2x_Ax_B)\ln(1-2x_Ax_B)
-
2x_Ax_B\ln(2x_Ax_B).
\end{equation*}
This expression is not equal, in general, to the entropy of mixing (eq. \eqref{eq:mixing_entropy}, which highlights that the correct entropy arises only from the conditional mapping entropy \(H(\Delta\mid X)\).

\section{Quantum Entropy}

For completeness, we briefly address the role of quantum information in the \texttt{asdf} framework when applied to real material systems. At $T = 0$ K, Clausius’s formulation \cite{Clausius1865-gi} of thermodynamic entropy asserts that $S = 0$. However, quantum systems retain nontrivial structure in their ground state, and the statistical description of such systems is naturally expressed through the von Neumann entropy\cite{Von_Neumann1996-tm}. In practice, the \texttt{asdf} method operates on atomic configurations extracted from molecular dynamics simulations (e.g., VASP), in which ionic motion is treated classically. As a result, quantum vibrational effects are not explicitly encoded in the information sources that are compressed within the \texttt{asdf} scheme.

To quantify the magnitude of these missing contributions, we begin with the von Neumann entropy,
\begin{equation*}
S = -k_B \mathrm{Tr}(\rho \ln \rho),
\end{equation*}
where $\rho = e^{-\beta \hat{H}} / Z$ is the canonical density operator and $Z = \mathrm{Tr}(e^{-\beta \hat{H}})$ is the partition function. Since
\[
\ln{\rho}=-\beta\hat{H}-\ln{Z}
\]
then
\begin{align*}
S&=-k_B\mathrm{Tr}\left[\rho(-\beta\hat{H}-\ln{Z})\right]
\nonumber\\
&=k_B\beta\,\mathrm{Tr}(\rho\hat{H})+k_B\ln{Z\,\mathrm{Tr}(\rho)}.
\end{align*}
Because $\mathrm{Tr}(\rho)=1$,
\[
S=k_B\beta\langle H\rangle + k_B \ln Z.
\]
Using
\[
\langle H\rangle=-\frac{\partial}{\partial\beta}\ln Z,
\]
this becomes
\begin{equation}
\label{eq:quantum_entropy}
S=k_B\left(\ln Z-\beta \frac{\partial \ln Z}{\partial \beta}\right).
\end{equation}
To model quantum vibrational effects, we approximate the ionic degrees of freedom as independent quantum harmonic oscillators. This approximation is well justified for solids near equilibrium, where lattice vibrations can be expressed in terms of normal modes (phonons), each of which behaves as a harmonic oscillator to leading order. The corresponding Hamiltonian for a single mode is
\begin{equation*}
\hat{H} = \hbar \omega \left( \hat{n} + \tfrac{1}{2} \right),
\end{equation*}
with eigenvalues
\begin{equation*}
E_n = \hbar \omega \left(n + \tfrac{1}{2}\right).
\end{equation*}

The partition function is therefore
\begin{equation}
\label{eq:quantum_partition}
Z = \sum_{n=0}^{\infty} e^{-\beta E_n}
= \frac{e^{-\beta \hbar \omega / 2}}{1 - e^{-\beta \hbar \omega}}.
\end{equation}

Substituting Eq. \eqref{eq:quantum_partition} into Eq. \eqref{eq:quantum_entropy} yields the quantum harmonic oscillator entropy,
\begin{equation*}
S_{\mathrm{QHO}} = k_B \left[
\frac{\beta \hbar \omega}{e^{\beta \hbar \omega} - 1}
- \ln \left(1 - e^{-\beta \hbar \omega}\right)
\right].
\end{equation*}

We can now compare this result with the classical harmonic oscillator entropy derived in Section~\ref{sec:classical_ho}. Evaluating both expressions for real material systems, as shown in Table~\ref{tab:quantum_classical_comparison}, reveals that the discrepancy between classical and quantum harmonic oscillator entropies is negligible at physically relevant temperatures. This is expected, as $\beta \hbar \omega \ll 1$ in this regime, suppressing quantum effects.

\begin{table*}[t]
\centering
\caption{Comparison of classical and quantum one-dimensional harmonic oscillator entropies for representative material systems. The discrepancy $\Delta S = S_{\mathrm{QHO}} - S_{\mathrm{CHO}}$ is reported along with its percentage relative to the total entropy of each phase. Entropy values are in $\mathrm{J}/\mathrm{K}/\mathrm{mol}$.}
\label{tab:quantum_classical_comparison}
\setlength{\tabcolsep}{4pt}
\begin{tabular}{lcccccccc}
\hline
Material & Phase & $T$ ($\mathrm{K}$) & $\Theta_D$\cite{Kittel} ($\mathrm{K}$) & $S_{\mathrm{CHO}}$ & $S_{\mathrm{QHO}}$ & $\Delta S$ & $S_{\mathrm{phase}}$\cite{NIST} & $\Delta S/S$ (\%) \\
\hline
Al & FCC  & 1000 & 428 & 15.370 & 15.434 & 0.064 & 62.25 & 0.103 \\
   & Liquid &      &     &        &        &       & 73.4  & 0.087 \\
\hline
Ti & BCC  & 1800 & 420 & 20.414 & 20.433 & 0.019 & 87.47 & 0.0217 \\
   & Liquid &      &     &        &        &       & 94.42 & 0.0201 \\
\hline
Si & Diamond  & 1400 & 645 & 14.758 & 14.831 & 0.073 & 56.46 & 0.129 \\
   & Liquid &      &     &        &        &       & 87.12 & 0.0838 \\
\hline
W  & BCC  & 3400 & 400 & 26.108 & 26.113 & 0.005 & 104.3 & 0.0048 \\
   & Liquid &      &     &        &        &       & 115.7 & 0.0043 \\
\hline
\end{tabular}
\end{table*}

Therefore, while the \texttt{asdf} method does not explicitly include quantum vibrational contributions, its classical treatment is sufficient for modeling thermodynamic entropy in real material systems at finite temperature.

For this analysis, Debye temperatures were taken from standard tabulated low-temperature limits \cite{Kittel}, which define the characteristic phonon energy scale; the results are insensitive to the precise value due to the high-temperature regime ($\Theta_D/T\ll 1$). Absolute entropy data for each material phase was taken from the NIST database \cite{NIST}.

\section{Generalization Beyond Analytically Solvable Hamiltonians}
\label{sec:generalization}

\paragraph{}
We first emphasize that the central identity of the \texttt{asdf} construction has already been proven in two analytically tractable limits: the ideal gas (IG) and the uncoupled one--dimensional harmonic oscillator (1D--HO). In both cases, we demonstrated that the entropy of a microstate $Y$ can be represented by the conditional Shannon entropy of a mapping $\Delta$ defined relative to an uncorrelated microstate $X$. Concretely, with $X$ drawn independently of $Y$, we showed that the conditional entropy of the induced map satisfies
\[
\label{eq:ig_ho_exact}
H(\Delta \mid X) = H(Y)
\]
for the IG and the 1D--HO. These results establish that the mapping formalism is exact in two extreme regimes: purely diffusive (gas-like) behavior and purely bound, approximately harmonic (solid-like) behavior.

\paragraph{}
A natural question is whether such agreement in limiting cases supports generalization to interacting, anharmonic Hamiltonians (e.g., liquids and solids with realistic many-body correlations). A useful perspective is provided by the two-phase thermodynamic (2PT) framework of Lin, Blanco, and Goddard\cite{LinBlancoGoddard2003}, which shows that the dynamical mode content of a liquid can be viewed as a superposition of gas-like (diffusive) and solid-like (non-diffusive) contributions. In 2PT, the vibrational density of states $S(\nu)$ extracted from a molecular dynamics trajectory is decomposed into a gas-like component (capturing diffusion through $S(0)\neq 0$) and a solid-like component (with $S(0)=0$), and thermodynamic properties are obtained by applying appropriate limiting-model statistics to each component. The key conceptual point is that a liquid is not treated as ``neither gas nor solid,'' but rather as a system whose dynamics contains both diffusive and vibrational signatures across frequencies, so that the thermodynamics can be reconstructed by combining these two characteristically distinct contributions.

\paragraph{}
This decomposition suggests a structural reason why a mapping-based entropy method that is exact in a diffusion-dominated limit (gas-like) and in a bound, harmonic limit (solid-like) may extend to liquids and other interacting systems: the relevant equilibrium dynamics empirically exhibits both behaviors. In this sense, the IG and HO represent canonical endpoints of the dynamical spectrum. The 2PT viewpoint therefore motivates the hypothesis that the \texttt{asdf} map $\Delta$ can generalize to more complex Hamiltonians by encoding (i) diffusion-like rearrangements and (ii) solid-like vibrational structure within a unified information-theoretic representation.

\subsection{$H(\Delta\mid X)$ as an approximation to $H(Y)$}
\label{subsec:approximation_gap}

\begin{figure}
    \centering
    \includegraphics[width=0.8\linewidth]{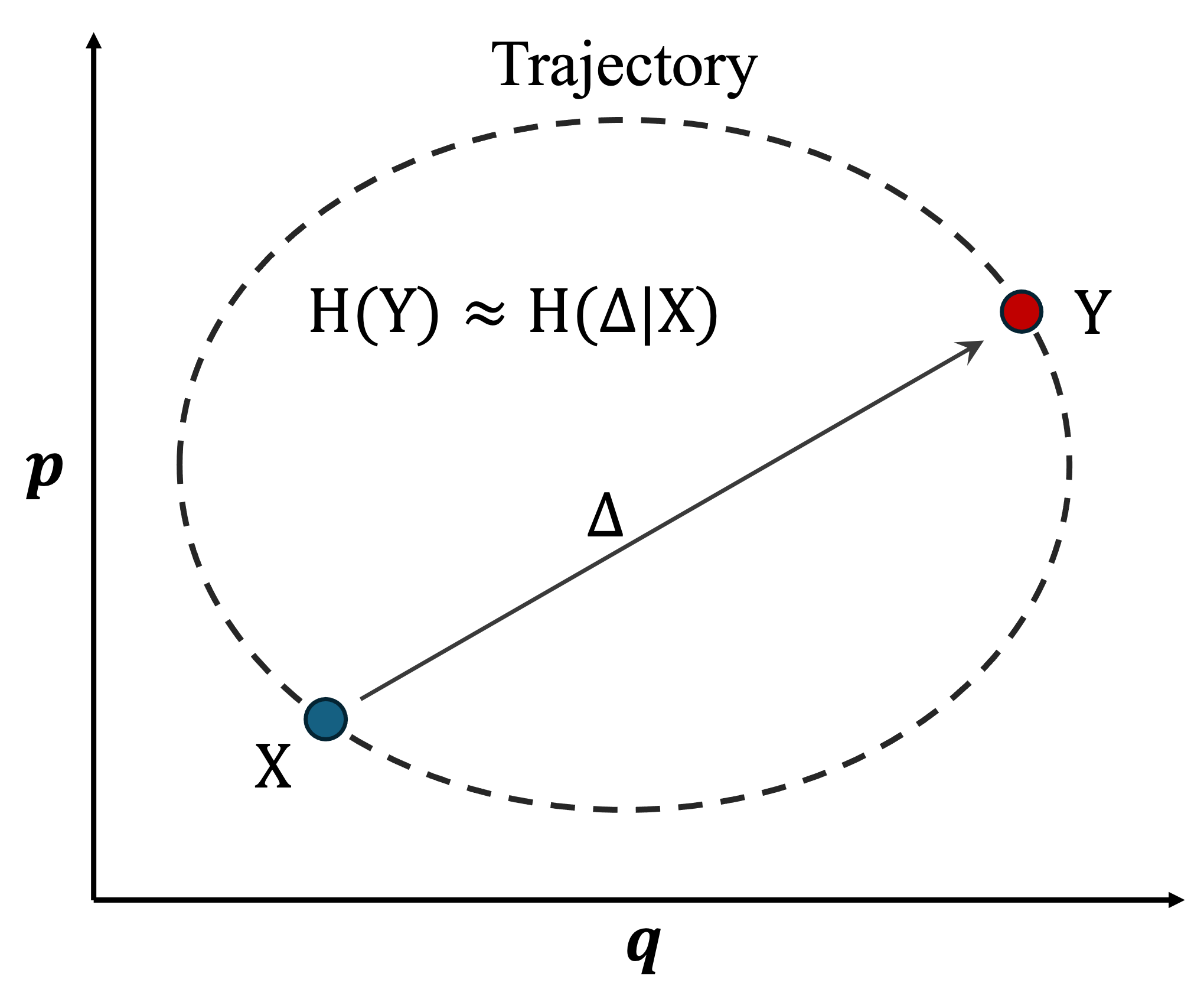}
    \caption{The entropy of a microstate $Y$ can be approximated as the conditional entropy of the mapping between an uncorrelated microstate $X$ and $Y$.}
    \label{fig:phase_space_arrows}
\end{figure}

\paragraph{}
To discuss general Hamiltonians, we make explicit the map used in practice. Let
\[
X=\{x_j\}_{j=1}^N,\qquad Y=\{y_i\}_{i=1}^N,
\]
be two configurations (microstates) of $N$ atoms in $\mathbb{R}^3$, with $X$ drawn independently of $Y$ from an ``uncorrelated'' reference distribution.
Define the nearest-neighbor assignment
\begin{equation}
\label{eq:sigma_def}
\sigma(i) \;=\; \arg\min_{j\in\{1,\dots,N\}} \|y_i-x_j\|,
\end{equation}
and the residual (displacement) vectors
\[
\label{eq:delta_def}
\delta_i \;=\; y_i - x_{\sigma(i)}.
\]
The mapping object is then
\[
\label{eq:Delta_def}
\Delta \;=\; \{\delta_i\}_{i=1}^N,
\]
which may be visualized as a set of vectors that ``emanate'' from anchor points in $X$ toward the atoms in $Y$ (see Fig.~\ref{fig:phase_space_arrows}).

\paragraph{}
The conditional entropy $H(\Delta\mid X)$ has an immediate operational meaning: it is the expected description length of $\Delta$ when both sender and receiver already know $X$. To see how this relates to $H(Y)$, imagine that the sender wishes to communicate $Y$ to a receiver who already possesses $X$. If the map were strictly one-to-one---i.e., if each anchor $x_j$ generated exactly one atom in $Y$---then the receiver could reconstruct $Y$ by simply applying $y_i = x_{\sigma(i)}+\delta_i$ under a fixed ordering convention, and $\Delta$ would (up to standard discretization) contain the full information needed to specify $Y$.

\begin{figure}
    \centering
    \includegraphics[width=0.5\linewidth]{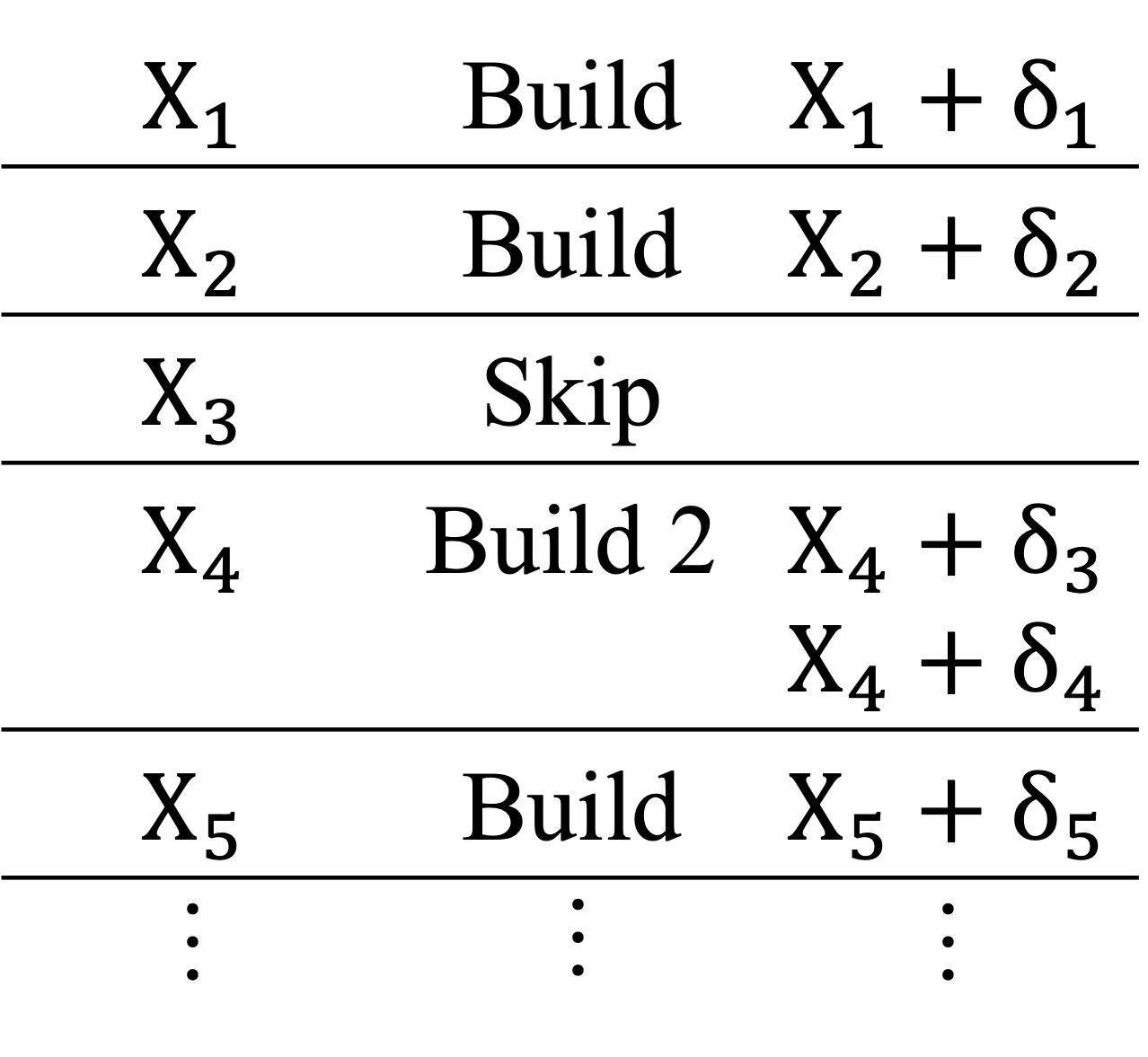}
    \caption{The information relayed from sender to receiver to construct a microstate $Y$ given a relative microstate $X$ can be constructed as procedural operations over a fixed sequence of the elements in $X$. For each element in $X$, information on how the receiver can build elements in $Y$ must be supplied.}
    \label{fig:instructions}
\end{figure}

However, for a general interacting Hamiltonian, the nearest-neighbor assignment in Eq.~\eqref{eq:sigma_def} need not be strictly one-to-one. In particular, multiple $y_i$ can share the same anchor $x_j$ (a ``double map''), and some anchors may produce no $y_i$ (a ``skip''). These events introduce additional \emph{discrete} information that is not contained in the continuous residual vectors alone unless it is encoded explicitly. Let $c_j$ denote the multiplicity (occupancy)

\[
\label{eq:counts}
c_j \;=\; \#\{\, i : \sigma(i)=j\,\},\qquad j=1,\dots,N,
\]
so that $\sum_{j=1}^N c_j = N$. In a fully decodable communication protocol, the sender would transmit, in addition to the residual vectors, the occupancy pattern $\{c_j\}$ indicating for each anchor $x_j$ whether the receiver should ``build'' $0$, $1$, $2$, $\dots$ atoms (see Fig.~\ref{fig:instructions}).

\paragraph{}
A mathematically clean way to formalize this is to define an \emph{augmented} code
\[
\label{eq:Delta_aug}
\Delta' \;=\; (\Delta, C),\qquad C=\{c_j\}_{j=1}^N.
\]
Given $X$ and $\Delta'$, one can reconstruct $Y$ by placing, for each $j$, exactly $c_j$ atoms at $x_j+\delta_{j,k}$ (with any fixed rule for grouping the relevant displacements). In this sense, $\Delta'$ is an information-complete description of $Y$ relative to $X$ (up to measure-zero degeneracies such as exact distance ties), so that in the idealized discrete setting
\[
\label{eq:invertible_code}
H(Y\mid X) = H(\Delta'\mid X).
\]
If $X$ and $Y$ are independent, then $H(Y\mid X)=H(Y)$, hence $H(\Delta'\mid X)=H(Y)$.

In practice, we do not transmit the full $C$ explicitly; instead, we approximate the occupancy statistics by assuming that deviations from the default $c_j=1$ are dominated by \emph{skips} ($c_j=0$) and \emph{doubles} ($c_j=2$), with $c_j\ge 3$ rare in condensed phases. This induces a \emph{cost} term corresponding to the additional bits needed to specify non-default occupancies:
\[
\label{eq:cost_term_general}
B_{\mathrm{cost}} \;\approx\; \sum_{j=1}^N L(c_j),
\]
where $L(c_j)$ is the code length assigned to the event ``$c_j$ differs from $1$'' under the chosen approximation (e.g., a binary flag for doubles and/or skips, with higher occupancies treated separately when they occur).

At the same time, a naive transmission of multiple residual vectors anchored at the same $x_j$ introduces an artificial ordering: one must list ``the first'' displacement and ``the second'' displacement, even though $Y$ consists of unlabeled atoms and the ordering of those two atoms is physically irrelevant. This produces an overcount which must be removed by an \emph{indistinguishability} (symmetry) savings term. The number of redundant orderings for an anchor with multiplicity $c_j$ is $c_j!$, hence the associated savings is
\[
\label{eq:savings_term}
B_{\mathrm{save}} \;=\; \sum_{j=1}^N \log_2(c_j!).
\]
Under the $\{0,1,2\}$ approximation, only doubles contribute since $\log_2(0!)=\log_2(1!)=0$ and $\log_2(2!)=1$.

\paragraph{}
Combining these observations yields a corrected estimator relating the mapping entropy to the microstate entropy:
\begin{equation}
\label{eq:corrected_entropy}
H(Y)\;\approx\; H(\Delta\mid X) \;+\; B_{\mathrm{cost}} \;-\; B_{\mathrm{save}},
\end{equation}
where $B_{\mathrm{cost}}$ accounts for missing multiplicity information (skips/doubles/rare higher occupancies), while $B_{\mathrm{save}}$ removes spurious information introduced by ordering indistinguishable atoms within a multi-occupancy anchor. Importantly, for condensed phases (solids and liquids) we observe that the dominant deviations from $c_j=1$ are indeed doubles (and occasionally triples), so that $B_{\mathrm{cost}}$ and $B_{\mathrm{save}}$ are comparable in magnitude and often nearly cancel. In such regimes, the leading contribution is therefore captured directly by $H(\Delta\mid X)$, consistent with the empirical success of the method in harmonic (solid-like) and diffusive (gas-like) limits.

Nevertheless, Eq.~\eqref{eq:corrected_entropy} provides a principled route to improved rigor when needed: one may explicitly compute the occupancy statistics $\{c_j\}$, include rare $c_j\ge 3$ events, and evaluate $B_{\mathrm{cost}}$ and $B_{\mathrm{save}}$ without approximation. This makes clear that the information in $\Delta$ conditioned on $X$ is not merely heuristic: when accompanied by the minimal discrete occupancy data (or an explicit correction), the mapping is sufficient to reconstruct $Y$ and thereby reproduce the microstate entropy in a controlled and systematically refinable manner.

\subsection{Momentum Contribution and Discretization Scale}
\label{subsec:momentum_discretization}

As noted prior, the \texttt{asdf} construction operates directly on the configurational degrees of freedom and does not explicitly encode momenta. In classical statistical mechanics, however, the partition function factorizes and yields a universal per atom momentum term (see Eq.~\ref{eq:S_ig_split})
\begin{equation}
\label{eq:Smom_standard}
S_{\mathrm{mom}} = \ln\!\left(\frac{1}{\Lambda^3}\right) + \frac{3}{2},
\end{equation}
where $\Lambda$ is the thermal de Broglie wavelength. This contribution depends only on temperature and particle mass, and is independent of the interaction potential.

Within the \texttt{asdf} framework, configurational entropy is computed as a Shannon entropy of discretized residual coordinates. For a continuous random variable $Z$ with differential entropy $h(Z)$, discretization at resolution $\epsilon$ produces a Shannon entropy\cite{information_dimension_original,information_dimension_ref}
\[
H_\epsilon(Z) = h(Z) + d\,\ln\!\left(\frac{1}{\epsilon}\right),
\]
where $d$ is the dimensionality. Thus, discretization introduces an additive term proportional to $\ln(1/\epsilon)$ per degree of freedom. When the threshold error $\epsilon$ is chosen to be of order the thermal de Broglie length $\Lambda$, the total per atom configurational discretization shift becomes
\[
\Delta S_{\mathrm{disc}} = 3 \ln\!\left(\frac{1}{\Lambda}\right),
\]
precisely reproducing the universal logarithmic part of the momentum entropy in Eq.~\eqref{eq:Smom_standard}.

In this sense, although momenta are not explicitly encoded in the mapping $\Delta$, their thermodynamic contribution emerges implicitly through the physically motivated discretization scale. The thermal de Broglie volume plays the role of the minimal phase-space cell, consistent with the standard quantum-mechanical resolution of classical Gibbs entropy. The appearance of the $3N\ln(1/\Lambda)$ term as a discretization correction is therefore not ad hoc but reflects the fact that thermodynamic entropy counts distinguishable phase-space cells. Choosing $\epsilon \sim \Lambda$ embeds the universal momentum contribution into the configurational compression scheme in a manner consistent with Eq.~\ref{eq:S_ig_split}.

This observation highlights an important structural point: the \texttt{asdf} method need only resolve configurational degrees of freedom, provided that the discretization scale is chosen in accordance with the natural quantum phase-space resolution. The momentum entropy, being universal and factorized, is thereby incorporated through the entropy measure itself rather than through explicit sampling of $\mathbf{p}$-space.

\subsection{Coarse Graining and Additional Degrees of Freedom}
\label{subsec:coarse_graining}

In realistic materials, the Hamiltonian generally contains additional internal degrees of freedom beyond nuclear coordinates and momenta. These may include electronic, magnetic, spin, or other internal variables. A general classical–statistical partition function can therefore be written schematically as \cite{Widom_2018_electronic}
\begin{equation}
Z = \int d\mathbf{p}\, d\mathbf{q}\, d\mathbf{s}\;
\exp\!\left[-\beta H(\mathbf{p},\mathbf{q},\mathbf{s})\right],
\label{eq:Z_general_more}
\end{equation}
where $\mathbf{p}$ and $\mathbf{q}$ denote nuclear momenta and positions, and $\mathbf{s}$ collectively denotes additional internal degrees of freedom (e.g.\ magnetic moments or electronic occupations). Under standard assumptions of separability, the Hamiltonian may be decomposed as
\begin{equation*}
H(\mathbf{p},\mathbf{q},\mathbf{s})
= K(\mathbf{p}) + V(\mathbf{q}) + H_{\mathrm{int}}(\mathbf{q},\mathbf{s}),
\end{equation*}
so that the partition function factorizes into momentum and non-momentum contributions. As discussed previously, the momentum integral produces a universal term that may be stored separately. The remaining configurational partition function becomes
\begin{equation*}
Z_{\mathrm{conf}} =
\int d\mathbf{q}\, d\mathbf{s}\;
\exp\!\left[-\beta \left(V(\mathbf{q}) + H_{\mathrm{int}}(\mathbf{q},\mathbf{s})\right)\right].
\end{equation*}

It is customary within statistical mechanics and materials thermodynamics to further factor this contribution according to physically distinct modes. In the absence of strong coupling between subsystems, the total entropy may be decomposed into additive components,
\begin{equation}
S = S_{\mathrm{config}} + S_{\mathrm{vib}} + S_{\mathrm{elec}} + S_{\mathrm{mag}},
\label{eq:S_decomposition}
\end{equation}
corresponding respectively to configurational disorder, vibrational motion, electronic excitations, and magnetic degrees of freedom. This separation is the standard coarse–graining paradigm\cite{VandeWalleCeder2002}: high-dimensional phase space is partitioned into subsystems whose entropic contributions may be evaluated independently to leading order. Existing methods work to approximate \eqref{eq:S_decomposition} through expansion into physically meaningful and/or obtainable PDFs \cite{WidomGao2019,Hong2025-ff}. The \texttt{asdf} construction developed thus far operates on nuclear configurations and therefore encodes the combined configurational and vibrational contributions,
\[
S_{\texttt{asdf}} \approx S_{\mathrm{config}} + S_{\mathrm{vib}},
\]
while the universal momentum term is accounted for separately as described earlier. 

For most non-magnetic metallic systems considered here, magnetic contributions are negligible, and the dominant additional correction arises from electronic excitations. Following the method of Widom \cite{Widom_2018_electronic}, the electronic entropy may be computed directly from the electronic density of states $D(E)$ obtained from first-principles calculations. At finite temperature, electronic occupations follow the Fermi–Dirac distribution
\begin{equation*}
f_T(E) = \frac{1}{\exp\!\left[(E-\mu)/k_B T\right] + 1},
\end{equation*}
and the corresponding electronic entropy is
\begin{align}
S_{\mathrm{elec}} =
&- k_B \int D(E)
[
f_T(E)\ln f_T(E)
\nonumber\\
&+ (1-f_T(E))\ln(1-f_T(E))
] dE.
\label{eq:S_elec_general}
\end{align}

This expression reduces at low temperature to the familiar linear form proportional to $D(E_F)T$, but at the elevated temperatures relevant to liquid metals the full integral in Eq.~\eqref{eq:S_elec_general} must be retained. 

The total entropy may therefore be constructed consistently as
\begin{equation}
S_{\mathrm{total}}
= S_{\texttt{asdf}} + S_{\mathrm{elec}},
\end{equation}
with magnetic terms omitted where appropriate. In this manner, the mapping-based configurational compression and the independently computed electronic density-of-states contribution combine within the standard coarse–grained thermodynamic framework to yield an entropy consistent with the generalized Hamiltonian in Eq.~\eqref{eq:Z_general_more}.

\subsection{Residual Mapping for Efficient Entropy Calculation}
\label{subsec:residual_mobility}

\begin{figure}
    \centering
    \includegraphics[width=1\linewidth]{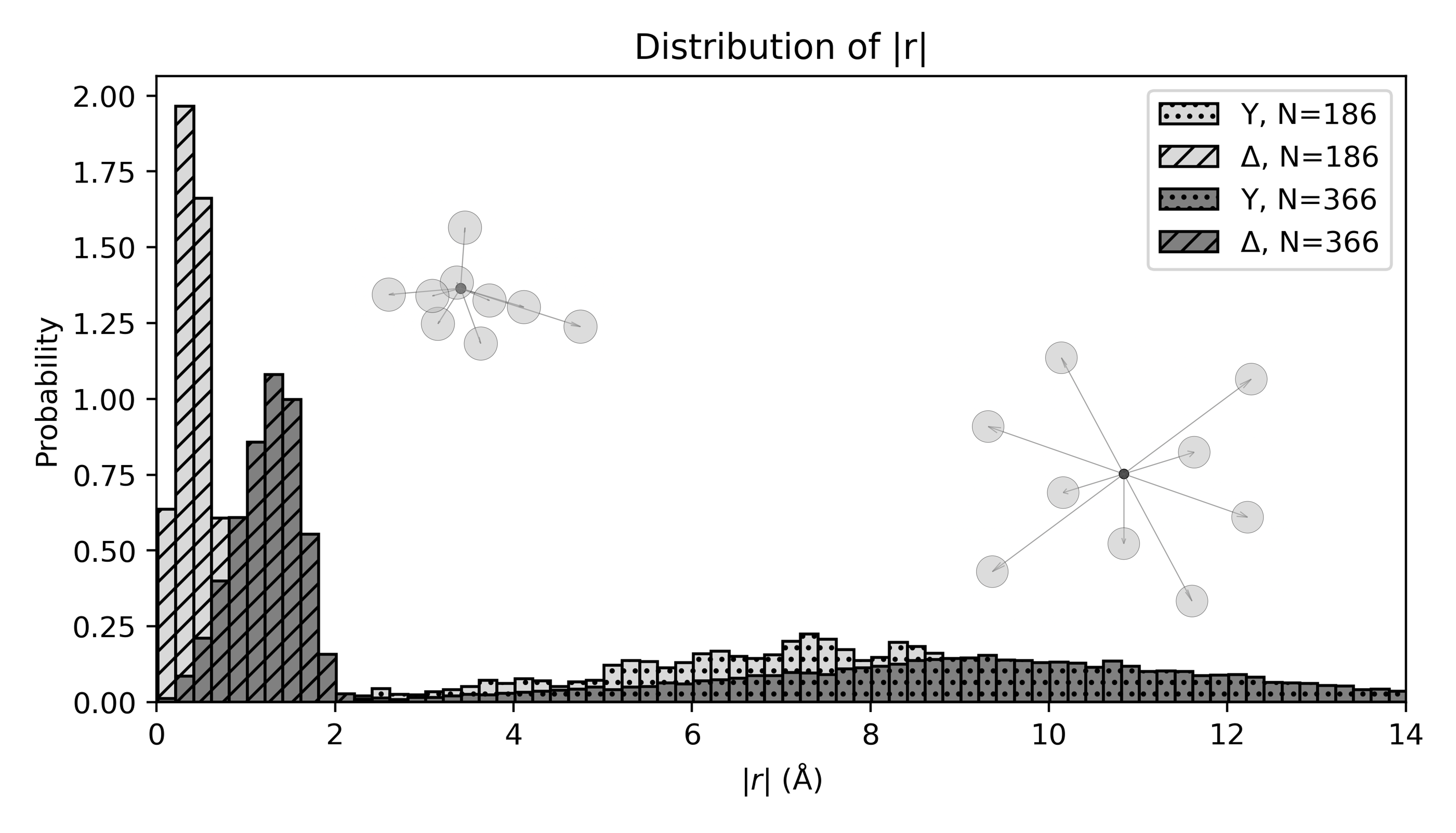}
    \caption{Distribution of vector magnitudes $r$ for raw and mapped ($\Delta$) data at varying system sizes for solid aluminum at $T=1000 \,\mathrm{K}$. The $\Delta$ representation produces a sharply peaked distributions near $r=0$ with reduced variance, reflecting a concentration of residual displacements that is favorable for finite-precision encoding.}
    \label{fig:N_N_distribution}
\end{figure}

The data shown in this section was generated from ab-initio molecular-dynamics simulations using the SLUSCHI\cite{SLUSCHI} package interfaced with VASP\cite{VASP1,VASP2,VASP3}. The Perdw-Burke-Ernzerhof \cite{PBE} exchange-correlation functional with the projector augmented-wave \cite{PAW} method was employed for high accuracy modeling of NPT ensembles. Simulated ensembles were then fed into \texttt{asdf} as outlined in our methodology paper\cite{asdf}, where the minimum mapping object ($\Delta$) is constructed and compressed to approximate the thermodynamic entropy from Eq. \ref{eq:asdf_entropy}.

The transformation from compressing a microstate $Y$ directly to compressing the residual map $\Delta(X,Y)$ reframes the entropy problem. Rather than estimating entropy through abstract enumeration of configurations, we quantify the information cost required to transform one typical configuration into another drawn from the same ensemble. In this representation, entropy measures state mobility. The relevant quantity is the conditional entropy $H(\Delta \mid X)$ introduced previously, which converts the problem from unconditional encoding of coordinates to conditional encoding relative to a structurally similar configuration.

The transformation therefore reduces the entropy problem to one of localized fluctuation encoding. In condensed phases, absolute coordinates span the full system scale $L$, whereas residual displacements are typically of vibrational magnitude $\ell \ll L$. 
The residual mapping concentrates probability mass near the origin of residual space, as seen in Fig.~\ref{fig:N_N_distribution}, and significantly reduces the dynamic range of the representation. While the total information content remains unchanged, this localization simplifies the statistical structure of the distribution and improves the numerical conditioning of compression algorithms as is implemented in the \texttt{asdf} framework. Fundamentally, $Y$ and $\Delta \mid X$ encode the same information, so they must have the same entropy. However, the Shannon entropy is the hard limit of compressibility, therefore any real implementation of an algorithm (such as \texttt{asdf}) will approximate the Shannon entropy as
\begin{equation}
\label{eq:shannon_is_limit}
    S_{\mathrm{limit}}=S_{\mathrm{algo}} + \delta_{\mathrm{algo}}
\end{equation}
where $S_{\mathrm{limit}}$ is the Shannon limit, $S_{\mathrm{algo}}$ is the algorithm dependent realized entropy, and $\delta_{\mathrm{algo}}$ is the algorithm-specific error, and performing transformations on the data set (such as $Y \rightarrow \Delta \mid X$) will conserve the true compressibility limit but may reduce the error in Eq. \eqref{eq:shannon_is_limit} \cite{elem_of_info_theory}.

\begin{figure}
    \centering
    \includegraphics[width=1.0\linewidth]{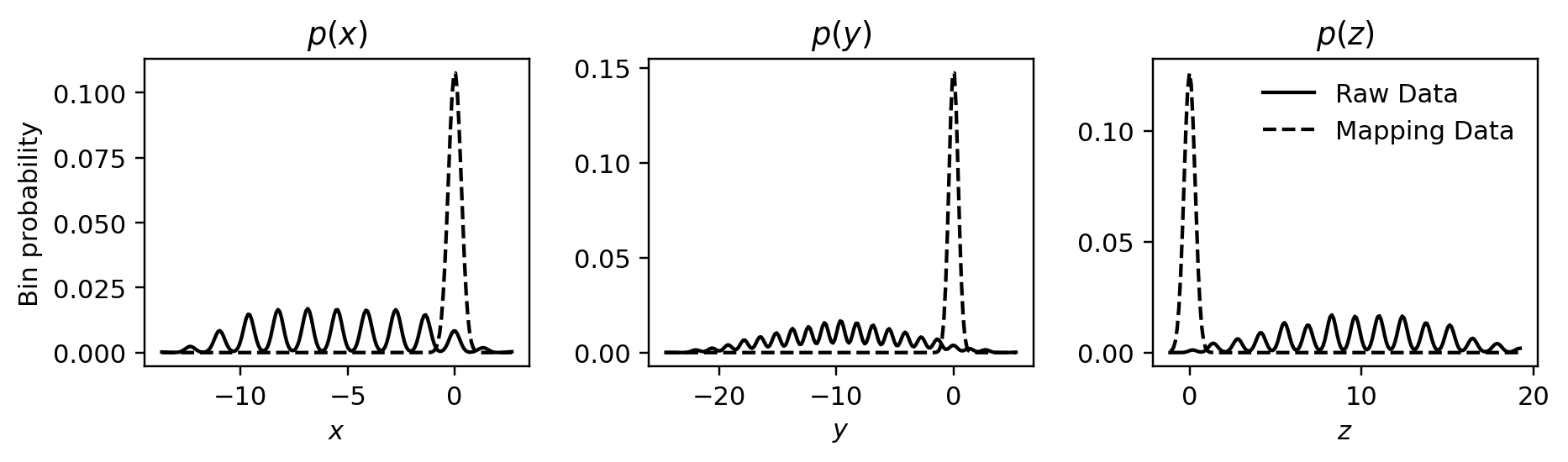}
    \caption{Marginal coordinate distributions $p(x)$, $p(y)$, and $p(z)$ for silicon in the diamond phase at $1400\,\mathrm{K}$, shown for both raw and mapped ($\Delta$) representations. The transformation from raw to $\Delta$ coordinates removes the lattice-scale positional structure and recasts the system in terms of local displacements about a reference configuration. As a result, the periodic features present in the raw coordinate distributions collapse into sharply localized peaks at zero in the $\Delta$ representation.}
    \label{fig:placeholder}
\end{figure}
An additional benefit arises from the suppression of superfluous information costs associated with numerical protocols. A simulated configuration may be decomposed schematically as
\[
Y = Y^{\mathrm{phys}} + \eta,
\]
where $\eta$ represents discretization artifacts and solver tolerances. For two configurations generated under the same protocol,
\[
\Delta = (Y^{\mathrm{phys}} - X^{\mathrm{phys}}) + (\eta_Y - \eta_X).
\label{eq:noise_cancel}
\]
Common-mode numerical contributions cancel to leading order in the difference. Direct compression of $Y$ would include entropy associated with $\eta$, whereas $H(\Delta \mid X)$ reduces sensitivity to representation-dependent artifacts and better isolates intrinsic ensemble fluctuations.

Thermodynamic entropy is inherently relative to a chosen phase-space measure and coarse-graining convention. Absolute entropy values are defined only up to additive constants. In direct coordinate representations, these baseline contributions must be managed explicitly. In the residual formulation, both configurations are drawn from the same stationary ensemble with identical discretization conventions. Ensemble-wide additive constants therefore cancel automatically, and no additional relative entropy construction is required. The relativity is internal to the ensemble. 

Another important advantage of the conditional mapping formulation is that it naturally resolves the ambiguity associated with absolute entropy baselines. In classical statistical mechanics, entropy is defined only up to an additive constant, reflecting the arbitrary choice of phase-space coarse-graining and reference measure. When entropy is estimated by directly compressing a microstate $Y$, this ambiguity manifests as a finite information cost required to specify absolute atomic positions and lattice vectors, even in a perfectly ordered ground state. As a result, such approaches do not trivially recover the thermodynamic requirement that $S \to 0$ as $T \to 0$ for systems with a unique ground state.
In contrast, the conditional mapping object $\Delta(X,Y)$ removes this baseline dependence by expressing entropy as relative information between microstates drawn from the same ensemble. At $0$ K, all microstates are identical (up to numerical representation), so the mapping produces $\delta_i = 0$ for all $i = 1,\dots,N$. The resulting distribution is a delta function at the origin, which when applied to equation \eqref{eq:shannon} as a discrete function with one bin of probability 1, yields a Shannon entropy of exactly 0. Therefore, evaluating entropy as $H(\Delta \mid X)$ yields $S=0$ without any explicit subtraction or calibration. This demonstrates that the mapping-based formulation intrinsically cancels additive constants associated with absolute coordinate descriptions and encodes only physically meaningful configurational variability, bringing it into direct correspondence with the thermodynamic definition of entropy.

The residual mapping also mitigates several practical difficulties that arise when working directly with absolute coordinates, particularly diffusion and periodic boundary conditions. In a conventional coordinate representation, atoms that cross a periodic boundary appear to undergo large discontinuous jumps in coordinate space, even though physically the motion is smooth. Such discontinuities complicate statistical analysis and compression because they introduce artificial large displacements that are unrelated to the local physics of the system. In the residual formulation, these artifacts are avoided by constructing the mapping relative to a reference configuration. When the mapping is defined, a periodic supercell of the reference configuration $X$ is constructed and the target microstate $Y$ is embedded within it. The nearest neighbor correspondence is then determined within this extended periodic representation. As a result, atoms that lie near opposite sides of the simulation cell are correctly associated with their nearest periodic images, ensuring that correlations across periodic boundaries are preserved and that the resulting residual vectors remain localized.

This construction also naturally accommodates diffusion, which presents a challenge for representations that rely on tracking individual atomic identities through time. In a diffusive system, atoms exchange positions and may migrate over long distances, making it difficult to define a consistent mapping between individual atoms in successive configurations. A representation based on fixed atom indices therefore introduces spurious large displacements whenever atoms exchange sites or cross boundaries. The residual mapping avoids this issue by defining the correspondence through a nearest-neighbor assignment rather than by preserving atom identity. Each atom in $Y$ is associated with the closest atom in the periodic supercell of $X$, producing a one-to-many mapping in which multiple atoms may be assigned to the same reference site when local rearrangements occur. Because the mapping depends only on spatial proximity, diffusive motion does not generate artificial long-range residual vectors, and the representation continues to capture only the physically relevant local displacements. It was shown in the Ideal Gas section that \texttt{asdf} successfully models the entropy of an ideal gas at reasonable temperatures, masses, and simulation sizes (Fig. \ref{fig:ideal_gas}). Since the ideal gas is diffusion dominant, the ability for \texttt{asdf} to reproduce the ideal gas entropy from analytic expressions implies \texttt{asdf} correctly handles diffusion.

Finally, the use of nearest-neighbor correspondence ensures that indistinguishability is treated consistently within the mapping procedure. Since atoms are not labeled by identity but rather associated through spatial proximity, permutations of identical atoms do not alter the encoded representation beyond the correction terms already introduced to account for multiple assignments. This approach therefore respects the indistinguishability of particles while maintaining a stable and localized description of configuration differences. By embedding the reference configuration in a periodic supercell and defining the mapping through nearest-neighbor correspondence, the residual representation simultaneously resolves issues associated with periodic boundary conditions, diffusive motion, and particle exchange, allowing the compression algorithm to operate on a representation that reflects the true local physics of the system.

Taken together, these results demonstrate that the residual transformation removes ensemble-wide structural redundancy, suppresses representation-dependent artifacts, eliminates additive baseline ambiguities, and remains invariant under physically relevant drift and diffusion processes. The entropy obtained from $H(\Delta \mid X)$ thus quantifies the intrinsic information associated with mobility within the ensemble, which is the physically meaningful content of thermodynamic disorder.

\section*{Declarations}
This research was supported by US Department of Defense Army Research Office Award number W911NF-23-2-0145, with use of Research Computing at Arizona State University. 

All data generated or analyzed during this study will be made publicly available upon publication of the manuscript.  
The ASDF codebase used in this study will be released in a public repository concurrent with publication.  
Instructions for reproducing entropy calculations and compression workflows will be provided alongside the code.
An online implementation of ASDF is currently available for interactive use at \url{https://faculty.engineering.asu.edu/hong/sluschi-api/}, allowing users to explore the method without requiring local installation.

The authors declare no competing interests. 

\bibliography{citations}

\end{document}